\documentclass[a4paper]{IEEEtran}
\usepackage[boxruled]{algorithm2e}
\usepackage{enumerate}
\usepackage{cite}
\usepackage{graphicx}
\usepackage{psfrag}
\usepackage{subfigure}
\usepackage{url}
\usepackage{amsmath}
\usepackage{amsthm}
\usepackage{array}
\usepackage{algorithm2e}
\usepackage{amssymb}
\usepackage{amsfonts}
\usepackage{float}

\newtheorem*{conjecture*}{Conjecture}
\newtheorem{lemma}{Lemma}

\newtheorem{remark}{Remark}
\newtheorem{definition}{Definition}
\newtheorem{theorem}{Theorem}
\newtheorem{example}{Example}

\title{An Improved Secretive Coded Caching Scheme exploiting Common Demands}
\begin{document}
\author{
}
\date{\today}
\author{
\IEEEauthorblockN{ Hari Hara Suthan C, Ishani Chugh and Prasad Krishnan\\}
\IEEEauthorblockA{Signal Processing and Communications Research Center,\\
International Institute of Information Technology, Hyderabad.\\
Email: \{hari.hara@research., chugh.ishani@research., prasad.krishnan@\}iiit.ac.in\\}
}
\maketitle
\thispagestyle{empty}	
\pagestyle{empty}
\begin{abstract}
Coded caching schemes on broadcast networks with user caches help to offload traffic from peak times to off-peak times by prefetching information from the server to the users during off-peak times and thus serving the users more efficiently during peak times using coded transmissions. We consider the problem of secretive coded caching which was proposed recently, in which a user should not be able to decode any information about any file that the user has not demanded. We propose a new secretive coded caching scheme which has a lower average rate compared to the existing state-of-the-art scheme, for the same memory available at the users. The proposed scheme is based on exploiting the presence of common demands between multiple users. 
\end{abstract}
\section{Introduction}
\label{sec1}
Caching has historically proved to be a significant aid in reducing the load in information flow networks, both wired and wireless.
The caching problem consists of two phases, the placement phase and the delivery phase. In the placement phase, when the network congestion is less, parts of files from the server are placed in the caches of the users(clients) based on statistics about the user demands. During the delivery phase when the users demand particular files, the network traffic is more, and the server ensures that the demands are met using network transmissions with the assistance of the cache. Caching is now almost ubiquitous in wireline and wireless networks, and newer variants are promised to accelerate the performance in latest communication infrastructures (see for example, \cite{EMM}).

Recently, in the landmark paper \cite{MaN}, a novel \textit{coded caching} scheme was proposed for a network consisting of a single server containing $N$ files each with $F$ bits connected via a single noiseless broadcast link to $K$ users. Each user is also equipped with a cache of size $M$ (which can store $MF$ bits).  Under this scenario, the authors of \cite{MaN} showed substantial reduction in the rate, i.e., the load on the shared link compared to the conventional caching scheme, by transmitting coded subfiles in the delivery phase.  
For instance, it was shown that for $N\geq K$, the uncoded conventional caching scheme achieves a rate of $K(1-\frac{M}{N})$, while the coded caching scheme achieves $\frac{K}{1+KM/N}(1-\frac{M}{N}).$ It was also shown that the rate achieved by the scheme of \cite{MaN} lies within a constant multiple of the optimal rate for that setup. 
Further refinements and extensions to this fundamental problem have been made since (for instance see \cite{ZLW,KNMD,YMA,TiZ}). The idea of exploiting commonality of demands between the users to minimise the average rate and the peak rate was presented in \cite{YMA}, where the authors showed the exact optimality of their scheme for uncoded cache placement in the original setup of \cite{MaN}. 

The problem of secretive coded caching was introduced in \cite{RPKP}. In secretive coded caching, the cache content and the transmissions are required to be such that each user can decode only the requested file by that user, and no information about other files. The general achievability scheme of \cite{RPKP} first encodes the files using a secret sharing scheme. Using the file-shares in the place of subfiles in the scheme of \cite{MaN}, and with added secret keys in the caching and delivery phase, the scheme of \cite{RPKP} ensures that information leakage of files to unintended users does not happen. A lower bound on the rate based on cut-sets is also derived in \cite{RPKP}. 
In \cite{ZeY1}, the authors consider a cache network where users are connected to the server via a set of relay nodes and present secretive coded caching schemes for such networks.  

In this work, we consider the problem of secretive coded caching and give an achievable scheme which has a lower average rate compared to the previous scheme in \cite{RPKP}. In order to present our scheme, we first analyse the leakage properties of a coded caching scheme modified from the general achievable scheme in \cite{RPKP} by removing the keys from the transmissions. The reason for analysing this `keyless' scheme is to understand which transmissions are redundant and can be removed from the scheme of \cite{RPKP} (thus decreasing the rate), without compromising on secrecy. Using our analysis, and based on insights from \cite{YMA}, we then propose a modified scheme which achieves an average rate better than \cite{RPKP} exploiting the commonality between the user demands.
The savings in the link utilization in our proposed scheme naturally are greater when the number of common demands are higher, and in the worst case of having no common demands at all, our scheme's link utilization matches that of \cite{RPKP}.

The rest of this paper is organized as follows. In Section \ref{sec2}, we recall the relevant results for secretive coded caching from \cite{RPKP}. We also give few definitions required for our purpose and summarise the main intuition behind this work. In Section \ref{sec3}, we analyse the `keyless' transmission scheme modified from that of \cite{RPKP}, while keeping the same cache content.  We derive the exact properties of the coded transmissions which result in leakage of file shares to unintended users, and also find the exact number of such leaked file shares. Using these results, in Section \ref{sec4}, we propose our secretive coded caching scheme based on a simple modification of the scheme in \cite{RPKP} and derive the improvements in average rate. 

\textit{Notations:} For a positive integer $n$, we denote by $[n]$ the set $\{1,...,n\}$. For some set $B$ and some $b\in B$, we denote by $B\backslash b$ the set $B\backslash\{b\}$. For some element $c$, we denote the set $B\cup \{c\}$ as $B\cup c$. Throughout the paper we assume that $\binom{n}{m}$ is zero, if $m>n$.
\section{Secretive Coded Caching}
\label{sec2}
We recall the problem setup and relevant results from \cite{RPKP}. As with \cite{MaN}, the cache network is a single server connected to the users via a broadcast link, with the parameters $N, K, M$ denoting the number of files (each of size $F$ bits) at the server, the number of users, and the cache available at each user. We denote the files at the server as $W_{k}: k\in [N]$, which are assumed to be independent random variables each distributed uniformly over $[2^F]$. The cache content at a user $k$ ($k\in[K]$), denoted by random variable $Z_k$, takes values from $[2^{MF}]$ according to some function of the $N$ files during the cache placement phase. The user demands $d_k:k\in[K]$ during the  delivery phase are collected in a $K$-length vector $\boldsymbol{d}$. The server then transmits a message $X_{\boldsymbol{d}} \in [2^{R_{\boldsymbol{d}}F}]$ which is a function of the files and the cache contents. The coded caching scheme is defined by the quantities $X_{\boldsymbol{d}}$ and $Z_k, k\in[K]$, and the quantity $R_{\boldsymbol{d}}$ is called the rate of the coded caching scheme given $\boldsymbol{d}$. Following \cite{YMA}, we denote by $N_e(\boldsymbol{d})$ the number of unique demands in $\boldsymbol{d}$. Let $U\subseteq[K]$ denote a set of \textit{leaders}, such that $|U|=N_e(\boldsymbol{d})$ and $d_{u_1}\neq d_{u_2}$ for any distinct $u_1,u_2\in U$. We also give the following definition for use in the present and forthcoming sections. 
\begin{definition}[Demand profile and Demand vector] Let $A$ be a subset of users, demanding $m$ unique demands ($m\leq |A|$)  in total. We define the \textit{demand profile} $d_p(A)$ of $A$ as the $m$-tuple of integers such that the $i^{th}$ value in the tuple denotes the number of users demanding the $i^{th}$ most requested file. We also define the \textit{demand vector of} $A$, denoted by $d(A)$, as the ordered $|A|$-tuple of demanded files by the users in $A$. We say that two demand vectors (of two sets $A,A'$) are equal (denoted by $d(A)=d(A')$) if they are equal as vectors except for a permutation. Note that $d([K])=\boldsymbol{d}$, the vector containing all the demands, which we shall refer to simply as the \textit{demand vector}.
\end{definition}

In \cite{RPKP}, the authors define the notion of \textit{information leakage} to quantify the amount of information that unintended users can decode information about files not demanded by it as follows. 
\begin{align*}
L=\underset{\boldsymbol{d}\in[N]^K}{\max} \underset{k\in [K]}{\max}~I(\boldsymbol{W}_{[N]\backslash d_k};X_{\boldsymbol{d}},Z_k),
\end{align*}
where $I(;)$ is mutual information, and $\boldsymbol{W}_{[N]\backslash d_k}$ is the set of files except for $W_{d_k}$.

We refer to a placement and delivery scheme such that  all users can recover their demands and $L=0$ as a \textit{perfectly secretive coded caching scheme}, or simply a secretive coded caching scheme. 
Assuming uniform distribution on the vector $\boldsymbol{d}\in[N]^K$, the average rate of a secretive coded caching scheme is defined as the expectation ${\mathbb E}_{\boldsymbol{d}}(R_{\boldsymbol{d}})$. 

The pair $(M,R_{avg})$ is said to be \textit{secretively achievable} if there exists a average rate $R_{avg}$ secretive coded caching scheme. In \cite{RPKP}, the authors presented a secretive coded caching scheme, which we refer to as the $SCC_{keys}$ scheme throughout this paper, achieving the below rate.
\begin{align}
\label{eqn101}
R_{\boldsymbol{d}}=R_{avg}^{keys}=
\begin{cases}
\frac{K(N+M-1)}{N+(K+1)(M-1)}, & \text{for}~M=\frac{Nt}{K-t}+1\\
1 & M=N(K-1),
\end{cases}	
\end{align}
where $t\in\{0,1,...,K-2\}$ and $R_{avg}^{keys}$ is the average rate of the scheme. The convex envelope of these memory-rate points for any general $1\leq M\leq N(K-1)$ is also shown to be achieved by memory sharing. 

We now present briefly the general achievability scheme from \cite{RPKP} (with rate as in (\ref{eqn101})) for $M$ values in the set $M=\{\frac{Nt}{K-t}+1: t=1,...,K-2\}$. We do not present the achievability scheme for $M\in \{1, N(K-1)\}$ given in \cite{RPKP} as it is not relevant to this work. 
The general achievability scheme given in \cite{RPKP} consists of a secret sharing outer code followed by a cache placement and delivery phase that follows the coded caching scheme of \cite{MaN} closely. We elaborate as follows.

A $(\binom{K-1}{t-1},\binom{K}{t})$ secret sharing scheme is first employed to encode each file $W_i$. From the file $W_i$, the secret sharing scheme generates  $\binom{K}{t}$ \textit{shares} (of size $F_s=\frac{F}{\binom{K}{t}-\binom{K-1}{t-1}}$ each) such that the file $W_i$ can be completely recovered from all the $\binom{K}{t}$ shares, but no information about $W_i$ is revealed by accessing any $\binom{K-1}{t-1}$ shares. The shares of $W_i$ are indexed by the $\binom{K}{t}$ subsets of $[K]$, denoted as $\{S_i^A:A\subseteq [K], |A|=t\}$. In the placement phase, for any user $k$ and any file $W_i$, the share $S_i^A$ is placed in the cache of $k$ if $k\in A$. In addition to shares, for each subset $\mathcal{A}\subset[K]$ such that $|\mathcal{A}|=t+1,$ an independent and uniformly generated key $T_{\mathcal{A}}$ of size $F_{s}$ bits is stored in cache at each user $k \in \mathcal{A}$. 
Note that for any user $k$, there are $\binom{K-1}{t-1}$ subsets of $[K]$ which are of size $t$ containing $k$. The memory occupied by the shares in the cache of any user is thus $N\binom{K-1}{t-1}F_{s}=F\frac{Nt}{K-t}$. Along with the keys, we thus get $MF=F(\frac{Nt}{K-t}+1)$, and hence $M=\frac{Nt}{K-t}+1$. 
During the delivery phase, the transmissions are as follows. For each $(t+1)$-sized subset ${\cal A}\subset[K]$, the vector $Y^{keys}_{\cal A}\triangleq T_{{\cal A}}+\sum_{x\in {\cal A}}S_{d_x}^{{\cal A}\backslash x}$ is transmitted. It is easy to check that the total number of bits transmitted is $R_{avg}^{keys}F$, where $R_{avg}^{keys}$ is as in (\ref{eqn101}). The decoding at any user $k$ is successful, as each missing share of a demanded file at $k$ is a summand in a transmission $Y^{keys}_{\cal A}$ for some $(t+1)$-sized subset ${\cal A}$ containing $k$. Any leakage of information from the cache is prevented by the secret sharing scheme, while leakage of information from the transmissions is prevented by the keys. For the purposes of this paper, the above presented coded caching scheme of \cite{RPKP} is referred to as the $SCC_{keys}$ scheme.
\subsection{Intuition behind this work}
In this paper, we present an improved perfectly secretive coded caching scheme for the same values of $M$ as in (\ref{eqn101}) other than $\{1,N(K-1)\}$. Our scheme has a lower average rate compared to (\ref{eqn101}). To do this, we exploit the presence of commonality between demands of different users. 
It is easy to notice that in the presence of all the demands being common, there is no coding required for the transmission. The entire scheme can simply be bypassed by transmitting the file which is being demanded as is, with secrecy being trivially maintained. It is therefore intuitively clear that the presence of common demands at the users can potentially aid in reducing the rate for those demands (and thus the average rate too). For the coded caching problem (without secrecy), this intuition was formalized in \cite{YMA}. 

In \cite{YMA}, it was shown that in the original scheme of \cite{MaN}, some transmissions are redundant, i.e., they can be obtained as linear combinations of other transmissions, for some instances of demand vectors. Thus, such transmissions can be `saved', i.e., they need not be transmitted and hence lead to reduced average rate.

The question we raise is - Can we `save' transmissions in the $SCC_{keys}$ scheme also? We notice that because of the presence of a unique key as a summand in each transmission of the $SCC_{keys}$ scheme, no transmission of $SCC_{keys}$ can obtained from  other transmissions. Thus there cannot be any further reduction in the average rate if we use the $SCC_{keys}$ scheme as it is. 

On the other hand, consider the minor modification to $SCC_{Keys}$. In this modified scheme, which we shall henceforth call as the $SCC_{keyless}$ scheme, the transmissions are $Y_{\cal A}=\sum_{x\in {\cal A}}S_{d_x}^{{\cal A}\backslash x}$ for each ${\cal A}\subseteq [K]$ such that $|{\cal A}|=t+1$. In other words, compared to the $SCC_{keys}$ transmission scheme, the keys are not included as summands in the $SCC_{keyless}$ scheme. The cache content for the $SCC_{keyless}$ scheme remains the same as with the $SCC_{keys}$ scheme. Clearly, except for using shares in the place of subfiles, this is identical to the original scheme of \cite{MaN}. Because of this similarity, as a direct consequence of Lemma 1 of \cite{YMA}, we have the following lemma. We leave the details of the proof to the reader.
\begin{lemma}
\label{lemmaYMA}
Let $\cal A$ be any $(t+1)$-sized subset of non-leaders from $[K]$ and ${\mathbb A}=\{{\cal A}_i:i=1,...,\binom{K}{t+1}-\binom{K-N_e(\boldsymbol{d})}{t+1}\}$ be the set of all $(t+1)$-sized subsets of $[K]$ such that ${\cal A}_i\cap U\neq \phi$. Then 
\begin{align*}
Y_{\cal A}=\sum_{\substack{{\cal A}_i\in{\mathbb A},d({\cal A}_i)=d({\cal A})\\ {\cal A}_i\backslash {\cal A}\subseteq U}}Y_{{\cal A}_i}
\end{align*}
\end{lemma}
Lemma \ref{lemmaYMA} suggests that the messages $Y_{\cal A}$ for any non-leader $(t+1)$-sized subset ${\cal A}$ need not be transmitted as it can be recovered from other transmissions, provided there is no violation of the secrecy constraint in the transmissions of $SCC_{keyless}$ (which depends on the demand vector $\boldsymbol{d}$). However, the following example shows that  $SCC_{keyless}$ ensures secrecy for some demand vectors, but results in leakage for others. Hence it is not a secretive scheme though the scheme lends itself to rate reduction by exploiting commonality of demands.

\begin{example}
\label{exm1}
Let $N=K=4$ and $t=2$. Here each file is encoded into $\binom{K}{t}=6$ shares. The number of transmissions required is $\binom{K}{t+1}=4$. Let the vector $\boldsymbol{d}=(1,1,2,2)$. Then the transmissions of $SCC_{keyless}$ scheme are
\begin{align*}
Y_{123}=S_{1}^{23}+S_{1}^{13}+S_{2}^{12},~~
Y_{124}=S_{1}^{24}+S_{1}^{14}+S_{2}^{12}\\
Y_{134}=S_{1}^{34}+S_{2}^{14}+S_{2}^{13},~~
Y_{234}=S_{1}^{34}+S_{2}^{24}+S_{2}^{23}
\end{align*}
For the sake of simplicity, in all examples in this paper, we drop the set notation in the subscript of the transmissions (for instance $Y_{\{1,2,3\}}$ is written as $Y_{123}$). It can be easily checked that there is no leakage at any user.  So $SCC_{keyless}$ performs as well as the $SCC_{keys}$ scheme in this case. Now, consider $\boldsymbol{d}=(1,1,1,2)$. Then the sum of the transmissions in the $SCC_{keyless}$ scheme 
$Y_{124}+Y_{134}+Y_{234}=S_{2}^{12}+S_{2}^{13}+S_{2}^{23}$. Hence, users $1,2,3$ can decode $S_{2}^{23}, S_{2}^{13}, S_{2}^{12}$ respectively. However, the $SCC_{keys}$ scheme remains secure for the same demand vector, with all the above transmissions having the keys as the extra summand.
\end{example}

In Section \ref{sec4} of this work, a new secretive coded caching scheme is proposed which can be said to combine the advantages of the $SCC_{keys}$ scheme (in ensuring secrecy for all demand vectors) and the $SCC_{keyless}$ scheme (in enabling rate reduction by not transmitting redundant transmissions). For this purpose, we investigate the leakage properties of $SCC_{keyless}$ scheme in Section \ref{sec3}. 

\section{Necessary and Sufficient Conditions for Leakage of Shares in $SCC_{keyless}$ scheme}
\label{sec3}

In this section, we obtain the precise leakage properties of the $SCC_{keyless}$ scheme. These properties will be used in the  description of our new improved secretive coded caching scheme in Section \ref{sec4}.

For some user $k$ and for a particular choice of demands at the $K$ users, we use the notation $E_k$ to denote the set of all users which have the same demand as $k$. 
The following lemma is an observation which will be used to show the main result in this section. 
\begin{lemma}
\label{onlyonecommon}
Any two distinct transmissions $Y_{X_1}$ and $Y_{X_2}$ of the $SCC_{keyless}$ (for some $(t+1)$-sized subsets $X_1,X_2$ of $[K]$) scheme have at most one summand share in common, i.e., at most one share is eliminated in the sum $Y_{X_1}+Y_{X_2}$. 
\end{lemma}
\begin{IEEEproof}
There is nothing to prove if there is no common share. Suppose there is a share common between $Y_{X_1}$ and $Y_{X_2}$. Then there must be some $x_1\in X_1,x_2\in X_2$ such that $x_1\neq x_2$ but $S_{d_{x_1}}^{X_1\backslash x_1}=S_{d_{x_2}}^{X_2\backslash x_2}.$ This means ${X_1\backslash x_1}={X_2\backslash x_2}.$ However this means that for any other $x_1'\neq x_1$ such that $x_1'\in X_1$, we have $X_1\backslash x_1'\not\subset X_2$. This concludes the proof.
\end{IEEEproof}
Before we give the main result in this section, we define the notion of a leaked share of a given coded caching scheme over a secret sharing scheme as an outer code.
\begin{definition}[Leaked shares]
\label{leakage}
Let ${\cal S}$ be a coded caching scheme where the cache and transmissions are functions of the shares (from the secret sharing scheme) and keys, and $X_{\boldsymbol{d}}$ be the set of transmissions of $\cal S$ for a particular choice of user demands $\boldsymbol{d}$. For some $t$-sized subset $X'$, a share $S_{n}^{X'}$ is said to be \textit{leaked} to some user $k$ from $X_{\boldsymbol{d}}$ if $n\neq d_k, ~k\notin X'$ and $S_{n}^{X'}$ is decodable from the set of transmissions $X_{\boldsymbol{d}}$. 
\end{definition}
The following theorem is the main result in this section which gives the necessary and sufficient conditions for the leakage of a share at a user.
\begin{theorem}
\label{leakageconditions}
A share $S_{n}^{X'}$ is leaked to a user $k$ in the $SCC_{keyless}$ scheme if and only if all the following conditions holds.
\begin{enumerate}
\item[C1]  $k\notin X'$, and there exists some user $x_1$ such that $n=d_{x_1}\neq d_k$.
\item[C2] The demand profile $d_p(X')=(t)$. 
\item[C3] Let $X_1=X'\cup x_1$. Then the demand profile $d_p(X_1)=(t,1)$.
\item[C4] Let $\{x_j:j=2,...,t+1\}=X'$. Then $X'\cup k\subseteq E_{x_2}$.
\end{enumerate}

\end{theorem}
\begin{IEEEproof}

\textit{If part:} As $X_1$ is a $(t+1)$-sized subset of $[K]$, the $SCC_{keyless}$ scheme has a well-defined transmission 
\begin{align}
\label{yx1}
Y_{X_1}=S_{d_{x_1}}^{X'}+S_{d_{x_2}}^{X_1\backslash x_2}+...+S_{d_{x_{t+1}}}^{X_1\backslash x_{t+1}}.
\end{align}
By the conditions given, we have a user $k$ such that $k\notin X_1$ but $k\in E_{x_2}$. To show that $S_n^{X'}$ is leaked to user $k$, it is sufficient to prove that there is a collection of transmissions whose sum results in a linear combination of the form 
\begin{align}
\label{eqn501}
S_{n}^{X'}+S_{n_1}^{X'_1}+S_{n_2}^{X'_2}+\hdots+S_{n_p}^{X'_p},
\end{align}
for some $p\geq 1$ such that $k\in X'_i,i=1,..,p$ and $n\neq d_k$ and $k\notin X'$ (by Definition \ref{leakage}).
We show that such a collection of transmissions does exist. Consider the set of transmissions $Y_{X_j}$, where $X_j=(X_1\backslash x_j)\cup k$, $j=2,...,t+1$. We write these transmissions explicitly as follows 
\begin{align*}
Y_{X_2}&=S_{d_{x_1}}^{X_2\backslash x_1}+S_{d_{k}}^{X_2\backslash k}+S_{d_{x_3}}^{X_2\backslash x_3}+...+S_{d_{x_{t+1}}}^{X_2\backslash x_{t+1}}\\
Y_{X_3}&=S_{d_{x_1}}^{X_3\backslash x_1}+S_{d_{x_2}}^{X_3\backslash x_2}+S_{d_{k}}^{X_3\backslash k}+...+S_{d_{x_{t+1}}}^{X_3\backslash x_{t+1}}\\
\vdots&=\vdots\\
Y_{X_{t+1}}&=S_{d_{x_1}}^{X_{t+1}\backslash x_1}+S_{d_{x_2}}^{X_{t+1}\backslash x_2}+...+S_{d_{x_t}}^{X_{t+1}\backslash x_t}+S_{d_{k}}^{X_{t+1}\backslash k}.
\end{align*} 
 We claim that the sum $\sum_{i=1}^{t+1}Y_{X_i}$ is of the form (\ref{eqn501}). To see this, firstly we note that $\forall j=2,...,t+1$, we have $X_j\backslash k=X_1\backslash x_j$ and $k \in X_j\backslash x_i, \forall  i\neq j$. Thus, all the shares whose index does not contain $k$ are eliminated in the sum $\sum_{i=1}^{t+1}Y_{X_i}$, except for $S_{d_{x_1}}^{X'}$ which is the share that is to be leaked. Furthermore there are $(t+1)^2-1$ shares (leaving out $S_{d_{x_1}}^{X'}$) totally considering all the transmissions $Y_{X_j},j=1,...,t+1$. By Lemma \ref{onlyonecommon}, at most $t(t+1)$ shares are eliminated in their sum (including the shares whose indices do not contain $k$). Thus, at least one share remains whose index contains $k$. Hence, the sum $\sum_{i=1}^{t+1}Y_{X_i}$ is of the form (\ref{eqn501}) with $p\geq 1$. This proves the if part.

\textit{Only if part:}
With respect to Condition C1, note that if $k\in X'$ then the share is already present in the cache of $k$. Also, if there is no user $x_1$ such that $n=d_{x_1}$, then the share $S_n^{X'}$ will not occur in any of the transmissions of $SCC_{keyless}$ scheme and there will thus be no possibility of its leakage. Finally if $n=d_k$, then leakage would be a misnomer as $W_n$ is intended for $k$. Hence Condition C1 holds.

We give the rest of the proof in three stages. 

\textit{Stage 1: Condition C1 holds, but Condition C2 doesn't hold:} 

Suppose there is a user $k$ at which $S_n^X$ is leaked. Then we must have $d_k\neq n$ and $k\notin X'$ (by Definition \ref{leakage}). Thus $k\notin X_1$. 

We now prove by contradiction. For leakage, there should be some linear combination of the transmissions such that (\ref{eqn501}) is satisfied. Since no direct transmission is of the form in (\ref{eqn501}) (as $k\notin X_1$ and $n\neq d_k$), at least two transmissions have to be linearly combined to get (\ref{eqn501}). Let a set of transmissions linearly combined to get (\ref{eqn501}) be denoted as ${\cal C}$. 

As Condition C1 holds, the transmission $Y_{X_1}$ is well defined as in (\ref{yx1}). We assume WLOG that the transmission $Y_{X_1}$ is such that in the linear combination of the transmissions in ${\cal C}$ leading to (\ref{eqn501}), the share $S_{n}^{X'}$ is not eliminated, but retained and thus leaked (clearly, such a transmission $Y_{X_1}$ must exist in ${\cal C}$). 

Since Condition C2 doesn't hold, we have $d_p(X')\neq (t)$, there must be some $y_1\in X'$ such that $y_1\notin E_k$.  Let $X=X_1\backslash\{x_1,y_1\}$. Then $Y_{X_1}$ can be written as 
\begin{align}
\label{eqn502}
Y_{X_1}=S_{d_{x_1}}^{y_1\cup X}+S_{d_{y_1}}^{x_1\cup X}+\sum_{x\in X}S_{d_x}^{X_1\backslash x}.
\end{align}
Note that, when $Y_{X_1}$ linearly combines with other transmissions in ${\cal C}$ to give (\ref{eqn501}), except for $S_{d_{x_1}}^{y_1\cup X}$, the other shares in (\ref{eqn502}) are necessarily eliminated because $k\notin x_1\cup X$ and also $k \notin X_1\backslash x$ for any $x\in X$. 

For the purposes of this proof, we henceforth denote a share $S_{d_{y}}^{x\cup X''}$ as $(d_{y},x,X'')$. In order to eliminate the share $(d_{y_1},x_1,X)$ in $Y_{X_1}$, we need a transmission $Y_{X_2}\in {\cal C}$, with $X_2=X\cup\{x_1,y_2\}$ with $d_{y_2}=d_{y_1}$. Obviously, we must have $y_2\neq y_1$ as otherwise $X_1=X_2$.

In $Y_{X_1}+Y_{X_2},$ the share $(d_{y_1},x_1,X)=(d_{y_2},x_1,X)$ gets eliminated, and no other share is eliminated as only at most one share is common between any two transmissions by Lemma \ref{onlyonecommon}. Thus, we now have at least one share $(d_{x_1},y_2,X)$ that has to be eliminated from $(Y_{X_1}+Y_{X_2})$ for leakage to occur at user $k$ (since $k\notin (y_2\cup X)$ as $k\notin E_{y_2}=E_{y_1}$).  For the sake of this proof, we call the share $(d_{x_1},y_2,X)$ the \textit{paired share} of the eliminated share $(d_{y_2},x_1,X)$ in $Y_{X_2}$. In order to eliminate this paired share $(d_{x_1},y_2,X)$, we must have another transmission $Y_{X_3}\in {\cal C}$ such that $X_3=X\cup\{x_2,y_2\}$ with $d_{x_2}\neq d_{x_1}$ but $x_2\neq x_1$. Clearly, the sets $X_1,X_2$, and $X_3$ are all distinct, and hence so are the transmissions $Y_{X_i}, i=1,2,3$. 

Note that the paired share of $(d_{x_2},y_2,X)$ is $(d_{y_2},x_2,X),$ and this has to be eliminated again. We continue the process of picking transmissions from ${\cal C}$ such that the paired share at every step is eliminated (the stopping criterion being that the paired share is eliminated by a prior picked transmission, thus not requiring us to pick a new transmission from ${\cal C}$). Because the number of transmissions is finite, the set ${\cal C}'\subseteq {\cal C}$ of transmissions including $Y_{X_1}$ picked to eliminate the paired shares is finite. 

Let the last-picked transmission $Y$ in ${\cal C}'$ be $Y_{X\cup\{x_r,y_{r'}\}}$ for some $r,r'$. Let the last but one transmission be $Y'$. Let $(d_{x_r},y_{r'},X)$ be the share eliminated in $Y$ by adding with the previous transmission $Y'$, and thus $(d_{y_{r'}},x_{r},X)$ is its paired share (the other possibility is that the $(d_{y_{r'}},x_r,X)$ is the eliminated share in $Y+Y'$, for which the proof proceeds similarly with only minor changes). 

We claim that the paired share $(d_{y_{r'}},x_{r},X)$ cannot be eliminated, thus contradicting the assumption that $Y$ is the last-picked transmission. The proof is as follows. Note that any transmission of ${\cal C}'$ prior to $Y$ is of the form $Y_{\{x_i,y_j\}\cup X}$ for some $x_i,y_j$. If $(d_{y_{r'}},x_{r},X)$ is to cancel with any share in some such prior transmission $Y_{\{x_i,y_j\}\cup X}$ of ${\cal C}'$ (which should have been picked prior to $Y'$ by Lemma \ref{onlyonecommon}), then it must be that $(d_{y_{r'}},x_{r},X)=(d_{y_{j}},x_{i},X)$ or $(d_{y_{r'}},x_{r},X)=(d_{x_{i}},y_{j},X)$. No other share of $Y_{\{x_i,y_j\}\cup X}$ can be equal to $(d_{y_{r'}},x_{r},X)$. 

Suppose $Y_{\{x_i,y_j\}\cup X}\neq Y_{X_1}$ and is picked prior to $Y'$. Then the shares $(d_{x_i},y_j,X)$ and $(d_{y_j},x_i,X)$ are eliminated already by adding with the two transmissions picked just prior to and just after $Y_{\{x_i,y_j\}\cup X}$. Furthermore, if $Y_{\{x_i,y_j\}\cup X} = Y_{X_1}$, $(d_{x_i},y_j,X)$ must not be eliminated (it is precisely the share which is leaked) while $(d_{y_j},x_i,X)$ is eliminated with $Y_{X_2}$. Thus no share is available in all transmissions prior to $Y'$ to eliminate $(d_{y_{r'}},x_{r},X)$, as all such shares are either eliminated already or must be preserved. This proves that ${\cal C}'$ cannot be finite. This proves that Condition C2 should be satisfied for leakage. 

\textit{Stage 2: Conditions C1,C2 hold, but not Condition C3}

Suppose $S_n^{X'}$ is leaked to a user $k$ from $Y_{X_1}$. We are given that $d_p(X_1)\neq (t,1)$ while $d_p(X')=(t)$. Thus, we must have $d_p(X_1)=(t+1)$, as this is the only other possibility. However since $n\neq d_k$, we thus have $\{x_1,y_1\}\in X_1$ such that $d_{x_1}=d_{y_1}\neq d_k$. The rest of the arguments for this stage follow that of Stage 1 (starting from the para with (\ref{eqn502})). This proves that Condition C3 should also hold.

\textit{Stage 3: Conditions C1, C2, C3 hold, but not Condition C4}

Note that $X'\subseteq E_{x_2}$ (by definition) and $k\notin X'$ (by Condition 1), thus the failure of Condition C4 means that $k\notin E_{x_2}$.

 Now consider $Y_{X_1}=Y_{X'\cup x_1}$ as in (\ref{eqn501}) from which $S_{d_{x_1}}^{X'}=S_n^{X'}$ is supposedly leaked to user $k$ with $d_{x_1}\neq d_k$. As $k\notin E_{x_2}$, there is at least one $y_1\in X'$ such that $y_1\neq x_1$ and $d_{y_1}\neq d_k$. Once again, we invoke the same arguments as Stage 1 (starting from the para with (\ref{eqn502})) to complete the proof of this stage,  showing that Condition 4 and hence the theorem should hold true.
\end{IEEEproof}
Using Theorem \ref{leakageconditions}, we now determine the set of all possible leaked shares in the $SCC_{keyless}$ scheme. 
\begin{lemma}
\label{leakedtransmissions}
Consider the transmissions of the $SCC_{keyless}$ scheme for a particular choice of demands at the users. The number of shares of some demanded file $W_n$ leaked to some user $k$ ($n\neq d_k$) is precisely $\binom{ |E_{k}|-1 }{t}$. Thus the user $k$ can decode $\binom{|E_k|-1 }{t}(N_e(\boldsymbol{d})-1)$ shares of the files not demanded by it. 
\end{lemma}
\begin{IEEEproof}
There is nothing to prove if $|E_k|\leq t$ or if $N_e(\boldsymbol{d})=1$. So we assume that $|E_k|\geq t+1$ and $N_e(\boldsymbol{d})\geq 2$. Let $x_1$ be some user such that $d_k\neq d_{x_1}$. Consider a subset $X'=\{x_2,..,x_{t+1}\}$ of $E_k$ such that $k\notin X'$. Let $X_1=X'\cup x_1$. The transmission $Y_{X_1}$ is well defined and contains the share $S_{d_{x_1}}^{X'}$ as a summand. Notice that all the four conditions of Theorem \ref{leakageconditions} are satisfied in this case. Thus $S_{d_{x_1}}^{X'}$ is leaked at user $k$. 

Note that we can pick set $X'\subset E_k$ (not containing $k$) in $\binom{|E_{k}|-1 }{t}$ ways. Each such $X'$ is unique, and thus so is the leaked share $S_{d_{x_1}}^{X'}$. Since there are $N_e(\boldsymbol{d})-1$ ways to choose $x_1$ (any user whose demand is not $d_k$ can be chosen), we thus have the total number of leaked shares at $k$ as $\binom{|E_{k}|-1 }{t}(N_e(\boldsymbol{d})-1).$ 

Finally, to show that no other share is leaked to $k$, we first note by the conditions of Theorem \ref{leakageconditions} that any leaked share must be of the form $S_n^{X'}$ where $X'\cup k \in E_k$ and $n\notin E_k$. Since we have already considered all such situations in the proof, no further leakage of shares is possible. This concludes the proof.

\end{IEEEproof}
\section{An Improved Secretive Coded Caching Scheme}
\label{sec4}
We now describe our improved secretive coded caching scheme, which we denote by $SCC_{common}$. The parameters $N,K$ and $t$ (and hence $M$) are as in Section \ref{sec2}. For a given vector of users demands $\boldsymbol{d}$, we have, as before, a set $U$ of leaders consisting of $N_e(\boldsymbol{d})$ users with all the unique demands. The cache placement phase remains the same as $SCC_{keys}$. 
After employing a $(\binom{K-1}{t-1},\binom{K}{t})$ secret sharing scheme to convert the files into shares as in $SCC_{keys}$ (with each share being of size $F_s=\frac{F}{\binom{K}{t}-\binom{K-1}{t-1}}$), the transmissions in the delivery phase are as follows.
\begin{itemize}
\item For each ${\cal A}\subseteq [K]$ of size $(t+1)$ such that demand profile $d_p({\cal A})=(t,1)$, transmit $Y^{keys}_{\cal A}=T_{{\cal A}}+\sum_{x\in {\cal A}}S_{d_x}^{{\cal A}\backslash x}$, where $T_{{\cal A}}$ is a independently generated key of size $F_s$ bits.
\item For each ${\cal A}\subseteq [K]$ of size $(t+1)$ such that ${\cal A}\cap U\neq \phi$ and $d_p({\cal A})\neq(t,1)$, transmit $Y_{\cal A}=\sum_{x\in {\cal A}}S_{d_x}^{{\cal A}\backslash x}.$ 
\end{itemize}

Before proving that the $SCC_{common}$ scheme is secretive and showing the improved average rate of the scheme, we first obtain the number of transmissions in the scheme. Note that in the scheme, transmissions are made corresponding to each $(t+1)$-sized subset ${\cal A}$ of $[K]$, except for those subsets of non-leaders (i.e ${\cal A}\subseteq [K]\backslash U$) with $d_p({\cal A})\neq (t,1)$. By abuse of terminology, we think of these sets of non-leaders with $d_p({\cal A})\neq (t,1)$ as corresponding to \textit{saved} transmissions, since corresponding to these sets also, transmissions are made in the $SCC_{keys}$ scheme. These saved transmissions translate to the reduced average rate of our scheme compared to $SCC_{keys}$. In order to calculate the number of saved transmissions, we partition the set of non-leaders $[K]\backslash U$ into demand classes $E_i':i=1,..,b$, such that the demands of users in each class is the same. We now have the following result on the number of saved transmissions. For the proof, we only have to count the number $(t+1)$-sized subsets ${\cal A}\subseteq [K]\backslash U$ with $d_p({\cal A})=(t,1)$. 
\begin{lemma}
\label{numberofsaved}
The number of saved transmissions in the $SCC_{common}$ scheme is $\binom{K-N_e({\boldsymbol{d}})}{t+1}-\Delta_t$, where $\Delta_t$ is as follows
\begin{itemize}
\item 
$\Delta_t=
\left(\sum\limits_{i=1}^{b}\binom{|E_i'|}{t}(K-N_e(\boldsymbol{d})-|E_i'|)\right)$,
if $t\geq 2$.
\item  If $t=1$, then 
$\Delta_t=\left(\sum\limits_{\{i,j\}\in {\cal D}}|E_i'||E_j'|\right),$
where ${\cal D}$ is the set of $\binom{b}{2}$ (unordered) pairs of elements from $[b]$.
\end{itemize}
\end{lemma}
\begin{IEEEproof}
Suppose $t\geq 2$. Then to construct a subset $\cal A$ as per our requirement, choosing $t$ users from $E_i'$ for any $i$ and the remaining one user from any of the other $(K-N_e(\boldsymbol{d})-|E_i'|)$ users, gives our result.

If $t=1$, then one user is chosen from any $E_i'$ (for some $i$) and the other from $E_j'$ (for some $i\neq j$) to obtain $\cal A$ as per our need. A standard counting argument for the number of ways completes the proof.
\end{IEEEproof}
We now give our main theorem which establishes the rate of our scheme and shows that it enables secrecy (using Theorem \ref{leakageconditions}) as well as correct decoding (using Lemma \ref{lemmaYMA}).
\begin{theorem}
\label{ratesavingthm}
The $SCC_{common}$ scheme is a secretive coded caching scheme and it achieves an average rate 
\begin{align*}
R_{avg}^{common}={\mathbb E}_{\boldsymbol{d}}\left(\frac{(\binom{K}{t+1}-\binom{K-N_e({\boldsymbol{d}})}{t+1}+\Delta_t)F_s}{F}\right),
\end{align*}
for $t=1,...,K-2$.
\end{theorem}
\begin{IEEEproof}
Firstly, we see that for successful decoding it is sufficient for the users to obtain transmissions either $Y_{\cal A}^{keys}$ or $Y_{\cal A}$ for each $(t+1)$-sized subset $\cal A$  of $[K]$. Now the only $(t+1)$-sized subsets for which neither $Y_{\cal A}$ nor $Y_{\cal A}^{keys}$ is available directly from the transmissions are those with $d_p({\cal A})\neq (t,1)$ and ${\cal A}\cap U=\phi$. However, by Lemma \ref{lemmaYMA}, any such $Y_{\cal A}$ is recoverable from transmissions $Y_{{\cal A}_i}$ (which are included in $SCC_{common}$) such that ${\cal A}_i\cap U\neq \phi$, $d({\cal A}_i)=d({\cal A})$, and ${\cal A}_i\backslash {\cal A}\subseteq U$. This is because such $Y_{{\cal A}_i}$s are transmitted in the $SCC_{common}$ scheme without having keys as summands. Thus, the decoding is successful at all users. Note that by Theorem \ref{leakageconditions}, for any transmission ${\cal A}$ of $SCC_{keyless}$ containing a leaked share, $d_p({\cal A})=(t,1)$. In $SCC_{common}$, any such transmission has an independently generated key as a summand, restricting the decoding of the summand shares to only intended users in ${\cal A}$. This completes the proof of secrecy. The achieved average rate is clear from Lemma \ref{numberofsaved} and the description of the scheme.
\end{IEEEproof}
\begin{remark}
The $SCC_{keys}$ scheme proposed by \cite{RPKP} has an average rate $\frac{\binom{K}{t+1}F_s}{F}$ (the result in (\ref{eqn101}) is obtained by simplifying this expression). Our scheme has a better average rate than $SCC_{keys}$ as $\Delta_t\leq \binom{K-N_e({\boldsymbol{d}})}{t+1}$. 
\end{remark}
	%
\begin{example}
\label{exm2}
Let $N=10$ and $K=10$ and $t=2$. Let the vector $\boldsymbol{d}=(1,1,1,1,1,2,2,2,2,3)$. 

Let us consider the set of leaders as $U = \{1,6,10\}$ and so $N_{e}(d)=3$. The sets of users with same demands are $E_{1}=\{1,2,3,4,5\}$, $E_{2}=\{6,7,8,9\}$, $E_{3}=\{10\}$. There will be $\binom{K}{t+1} = 120$ transmissions in the $SCC_{keys}$ scheme, and all the transmissions are with keys. 
Now, to calculate the number of transmissions in $SCC_{common}$ scheme, we partition the non-leader set $\{2,3,4,5,7,8,9\}$ into the demand classes $E_{i}', i=1,2,3$ with users with the same demand in each class, and obtain $E_{1}'=\{2,3,4,5\}$, $E_{2}'=\{7,8,9\}, E_3'=\phi$. Using Lemma \ref{numberofsaved}, we obtain $\Delta_t= \left(\binom{4}{2}(10-3-4) + \binom{3}{2}(10-3-3)\right) = 30$, while $\binom{K-N_{e}(d)}{t+1} = 35$. Thus the number of saved transmissions compared to the $SCC_{keys}$ scheme is $5$, and the $SCC_{common}$ scheme requires $115$ transmissions. The transmissions of $SCC_{common}$ with keys correspond to all the subsets of $[K]$ of size $(t+1)=3$ with demand profile $(t,1)=(2,1)$. It is a simple counting argument to show that this is equal to $\binom{5}{2}(5)+\binom{4}{2}(6)=86$. The remaining $29$ transmissions are sent without keys. The rate $R^{common}$ corresponding to this choice of demands in the $SCC_{common}$ scheme is easily seen to be approximately $0.96R_{avg}^{keys}$. 

\end{example}
\section*{Acknowledgment} This work was supported partly by the Early Career Research Award (ECR/2016/000447) from Science and Engineering Research Board (SERB) to Prasad Krishnan. Hari Hara Suthan was supported by the Visvesvaraya  PhD scheme for Electronics and IT.

\end{document}